\documentclass[%
 reprint,
 amsmath,amssymb,
 aps,
 superscriptaddress,
 showkeys,
prb,
floatfix,
]{revtex4-2}

\usepackage{graphicx}% Include figure files
\usepackage{dcolumn}% Align table columns on decimal point
\usepackage{bm}% bold math
\usepackage{hyperref}% add hypertext capabilities
\hypersetup{colorlinks, allcolors=blue}
\usepackage{physics}

\usepackage{siunitx}
\DeclareSIUnit\rydberg{Ry}
\begin{document}

%\title{Strain- and temperature-enhancement of unconventional charge density wave in ScV$_6$Sn$_6$ kagome metal}
%\title{Enhanced charge density wave in ScV$_6$Sn$_6$ kagome metal via strain and temperature control}
\title{Strain-induced enhancement of the charge-density-wave in the kagome metal ScV$_6$Sn$_6$}

\author{Manuel Tuniz**}
\affiliation{Dipartimento di Fisica, Universita degli studi di Trieste, 34127, Trieste, Italy}

\author{Armando Consiglio**}
\affiliation{Institut f\"{u}r Theoretische Physik und Astrophysik and W\"{u}rzburg-Dresden Cluster of Excellence ct.qmat, Universit\"{a}t W\"{u}rzburg, 97074 W\"{u}rzburg, Germany}
\affiliation{Istituto Officina dei Materiali, Consiglio Nazionale delle Ricerche, Trieste I-34149, Italy}

\author{Ganesh Pokharel}
\affiliation{Materials Department, University of California Santa Barbara, Santa Barbara, California 93106, USA}

\author{Fulvio Parmigiani}
\affiliation{Dipartimento di Fisica, Universita degli studi di Trieste, 34127, Trieste, Italy}
\affiliation{Elettra - Sincrotrone Trieste S.C.p.A., Strada Statale 14, km 163.5, Trieste, Italy}

\author{Titus Neupert}
\affiliation{Physik-Institut, Universit\"at Z\"urich, Winterthurerstrasse 190, CH-8057 Z\"urich, Switzerland}

\author{Ronny Thomale}
\affiliation{Institut f\"{u}r Theoretische Physik und Astrophysik and W\"{u}rzburg-Dresden Cluster of Excellence ct.qmat, Universit\"{a}t W\"{u}rzburg, 97074 W\"{u}rzburg, Germany}

\author{Giorgio Sangiovanni}%\email{sangiovanni@physik.uni-wuerzburg.de}
\affiliation{Institut f\"{u}r Theoretische Physik und Astrophysik and W\"{u}rzburg-Dresden Cluster of Excellence ct.qmat, Universit\"{a}t W\"{u}rzburg, 97074 W\"{u}rzburg, Germany}

\author{Stephen D. Wilson}
\affiliation{Materials Department, University of California Santa Barbara, Santa Barbara, California 93106, USA}

\author{Ivana Vobornik}
\affiliation{Istituto Officina dei Materiali, Consiglio Nazionale delle Ricerche, Trieste I-34149, Italy}

\author{Federico Salvador}
\affiliation{Istituto Officina dei Materiali, Consiglio Nazionale delle Ricerche, Trieste I-34149, Italy}

\author{Federico Cilento}\email{federico.cilento@elettra.eu}
\affiliation{Elettra - Sincrotrone Trieste S.C.p.A., Strada Statale 14, km 163.5, Trieste, Italy}

\author{Domenico Di Sante}\email{domenico.disante@unibo.it}
\affiliation{Department of Physics and Astronomy, University of Bologna, 40127 Bologna, Italy}
%\affiliation{Center for Computational Quantum Physics, Flatiron Institute, 162 5th Avenue, New York, NY 10010, USA}

\author{Federico Mazzola}\email{federico.mazzola@unive.it}
\affiliation{Istituto Officina dei Materiali, Consiglio Nazionale delle Ricerche, Trieste I-34149, Italy}
\affiliation{Department of Molecular Sciences and Nanosystems, Ca’ Foscari University of Venice, 30172 Venice, Italy}

\date{\today}

\begin{abstract}
The kagome geometry is an example of frustrated configuration in which rich physics takes place, including the emergence of superconductivity and charge density wave (CDW). Among the kagome metals, ScV$_6$Sn$_6$ hosts an unconventional CDW, with its electronic order showing a different periodicity than that of the phonon which generates it. In this material, a CDW-softened flat phonon band has a second-order collapse at the same time that the first order transition occurs. This phonon band originates from the out-of-plane vibrations of the Sc and Sn atoms, and it is at the base of the electron-phonon-coupling driven CDW phase of ScV$_6$Sn$_6$. Here, we use uniaxial strain to tune the frequency of the flat phonon band, tracking the strain evolution via time-resolved optical spectroscopy and first-principles calculations. Our findings emphasize the capability to induce an enhancement of the unconventional CDW properties in ScV$_6$Sn$_6$ kagome metal through control of strain.
\end{abstract}

\maketitle
** These authors contributed equally\\

A CDW is an emergent electronic phenomenon that manifests itself with standing waves and electronic reconstructions \cite{Gruner_1988, Carpinelli_1996}. This phenomenon typically involves a lattice distortion, with the mediation of electron-phonon coupling facilitating the phase transition. However, CDWs can arise from various driving mechanisms, also including electron-electron interactions, hence, in general, both electron-phonon coupling and electron-electron interactions contribute to their emergence ~\cite{Rossnagel_2011,Monceau_2012, doi:10.1080/23746149.2017.1343098}. In this respect, kagome metals are intriguing candidates to study the interplay of electrons and phonons, because they are characterized by a rich phase diagram with coexisting many-body phases such as superconductivity and CDW~\cite{Teng_2022, Ortiz2020, Luo_2022, PhysRevB.105.165146, Nie_2022, lin2024giant, Liang_2021}. Among kagome metals, ScV$_6$Sn$_6$ is a notable example, with an unconventional CDW arising from a bosonic mode collapse, driven by the softening of a flat-phonon band~\cite{Tuniz2023, Korshunov_2023, Arachchige_2022, Cao_2023, Lee_2024}.

\begin{figure*}
\centering
\includegraphics[width=\textwidth,angle=0,clip=true]{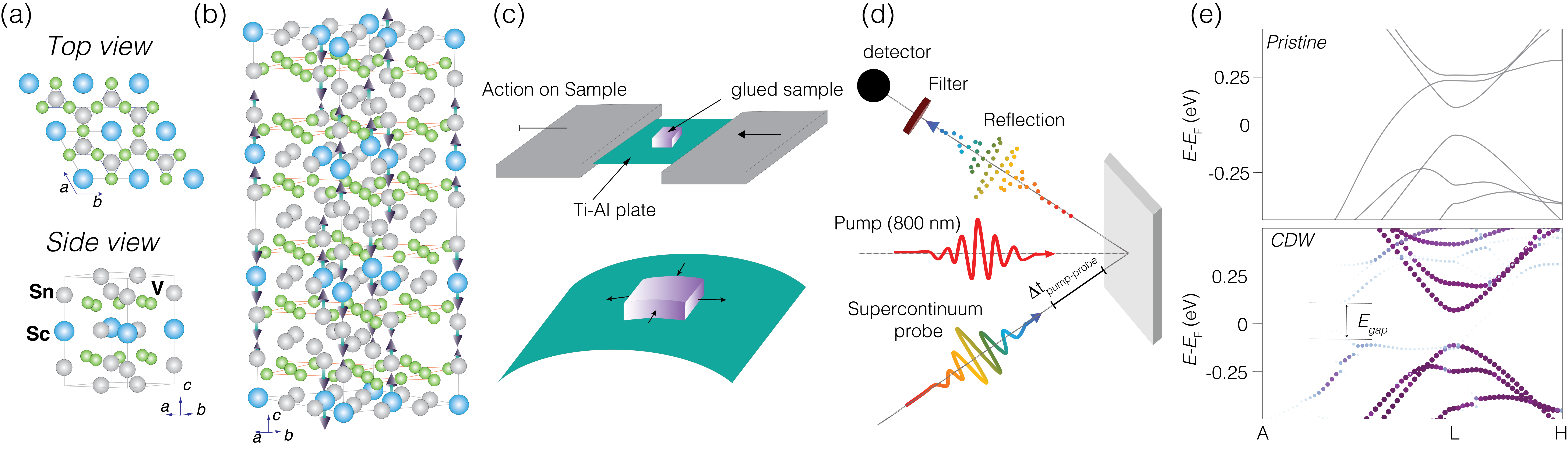}
\caption{(a) Top view and side view of the crystal structure of ScV$_6$Sn$_6$ in the absence of CDW. The vibrations which lead to the CDW transitions involve the Sc and Sn atoms (blue and gray atoms, respectively). (b) Crystal structure of ScV$_6$Sn$_6$ in the CDW phase. The unit cell gets significantly larger to accommodate the nonequivalent atoms. (c) Schematic of the strain device. The device consists of two blocks on which a clamp is attached. A block remains fixed and has a hole with a thread (screw thread). The other block, which contains the screws, is mechanically activated and is able to deliver strain. In between the clamps (see below), a Ti-Al plate is positioned and it gets squeezed by the strain. On such a plate, the sample is glued with silver epoxy. The strain is eventually delivered to the sample. A side view is also shown, focusing on the bending mechanism which delivers the strain. (d) Sketch of the time-resolved reflectivity experimental setup. The reflected probe beam is filtered and detected by an InGaAs photodiode detector. (e) DFT electronic structure along the A-L-H high-symmetry path, focused on the gap region above (upper panel) and below (lower panel) the transition temperature. The calculations are presented for a strain level of -2$\%$.}
\label{figure1}
\end{figure*}

ScV$_6$Sn$_6$ is a bilayer kagome metal, with an electronic bandstructure that comprises graphene-like Dirac states and flat bands typical of the kagome metal family \cite{Kim_2023, Kang_2020,Kang_2020b, Kang_2023}. Recently, it has also been shown to host topological properties \cite{Disante_2023}. ScV$_6$Sn$_6$ is the only known member of the \textit{so-called} 166-family of kagome bilayers which undergoes a CDW transition (the lattice structure is shown in Fig.~\ref{figure1}(a) above and in Fig.~\ref{figure1}(b) below the critical transition temperature), whose origin has been attributed to phonons and electron-phonon coupling, and specifically to the out-of-plane vibrations of Sc and Sn atoms~\cite{Lee_2024, Tuniz2023, Gu_2023}. The resulting charge order has been described as unconventional because the phonon $\mathbf{q}$-vector which generates the modulation is different from the order itself \cite{Korshunov_2023, Cheng_2024, Pokharel_2023}. The mechanism of this transition is also unusual, and attributed to the softening of a flat phonon band which, contrary to the first order nature of the transition, is a second order effect~\cite{Lee_2024, Cao_2023}. The softening is controlled by the electron band geometry and non-trivial quantum metric contributions to the electron-phonon coupling \cite{Yu_2023}. Moreover, the low-energy longitudinal phonon with propagation vector $\mathbf{q}=(1/3, 1/3, 1/2)$ collapses without the emergence of a corresponding long-range order. Rather, a long-range order occurs with a propagation vector $\mathbf{q}_{\text{CDW}}=(1/3, 1/3, 1/3)$~\cite{Korshunov_2023}.
As such, tuning the phonon mode is of considerable importance to control topological aspects in electron-phonon-coupled systems. We use ScV$_6$Sn$_6$ as the ideal material platform.

Previous studies from optical spectroscopy and time-resolved reflectivity described how the CDW transition occurs and the way the CDW can be quenched upon intense photoexcitation \cite{Tuniz2023, KimIR_2023}. In addition, photoelectron experiments evidenced the opening of a large energy gap ($\approx$200 meV) below the critical temperature of 92 K and such a gap manifests mostly along the $\bar{\Gamma}$-$\bar{\text{M}}$ high-symmetry direction of the Brillouin zone~\cite{Lee_2024}. Time-resolved optical spectroscopy (TR-OS) brought evidence that the lattice degrees of freedom play the prominent role in the CDW stabilization~\cite{Tuniz2023}, with the amplitude mode of the CDW displaying high resilience against intense photoexcitation. 
In this Letter, time-resolved optical spectroscopy (TR-OS) is used to detect the frequency of the amplitude mode (AM) of the CDW as a function of the applied uniaxial strain. By combining the outcome of TR-OS with density functional theory (DFT) calculations, we not only show that strain is an effective tool for controlling the CDW in ScV$_6$Sn$_6$, but we demonstrate that strain enhances the CDW, since the AM frequency is found to increase upon strain application. In detail, we show that at low temperature the effect of compressive strain on the phonon frequency is negligible, whereas on approaching the transition temperature, the phonon frequency is markedly increased by strain.
%, that eventually contrasts the temperature-induced softening of the AM  

We first describe the experimental setup used to apply strain. Uniaxial compression is delivered to the material through mechanical pressure application (see Fig.~\ref{figure1}(c)), similarly to methodologies reported in previous studies~\cite{Nicholson_2021, Sunko_2019}. The sample is glued on a flexible Ti-Al plate (more than 90\% Ti and less than 10\% Al). Such a plate is secured in between two clamps, which sit on a standard sample holder. One of the clamp has a fixed position, the other can be moved mechanically through a screw. When a compression is applied, the plate, on which the crystal is mounted, bends and delivers strain to the samples (see lower panel of Fig.~\ref{figure1}(c)). This device, simple but effective, allows us to perform measurements with various amount of strain. In order to quantify the actual strain transferred from the plate to the sample, which has a robust 3D structure, we rely on the result of our DFT calculations. In order to detect the effect of strain on the CDW, we make use of TR-OS, performed at the T-ReX laboratory (FERMI, in Trieste). The measurements were performed using a Ti:sapphire femtosecond (fs) laser system, delivering, at a repetition rate of 250~kHz, $\sim 50$~fs light pulses at a wavelength of 800~nm (1.55~eV). Time-resolved reflectivity experiments were performed at a probe wavelength of 1300~nm, obtained by filtering a broadband (550-1600~nm) supercontinuum probe beam, generated using a sapphire window. A cartoon of the TR-OS setup is shown in Fig.\ref{figure1}d.

\begin{figure*}
\centering
\includegraphics[width=\textwidth,angle=0,clip=true]{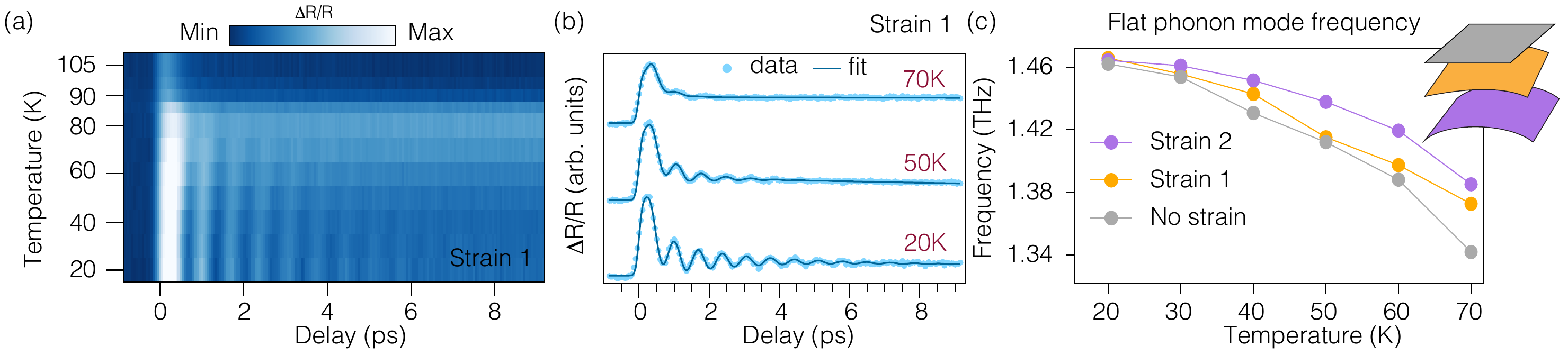}
\caption{(a) Two-dimensional map showing the evolution of the $\Delta R/R$ signal of a strained ScV$_6$Sn$_6$, as a function of the temperature and the pump-probe delay. The $\Delta R/R$ signal displays a strong coherent oscillation which adds up to an incoherent, exponentially decaying signal. (b) Traces extracted from (a) showing the modulation of the $\Delta R/R$ signal arising from the excitation of the CDW amplitude mode, for three selected temperatures below the CDW transition temperature. (c) Evolution of the frequency of the CDW amplitude mode as a function of temperature for three different amounts of compressive strain. The error bars estimated by our fitting procedure stay within the size of the dots.}
\label{figure2}
\end{figure*}

In Fig.~\ref{figure2}, we show the evolution of the normalized time-resolved reflectivity variation ($\Delta R/R$) as a function of temperature, for one strained configuration. The $\Delta R/R$ signal (see Fig.~\ref{figure2}a) is composed of two components: a coherent and an incoherent response. The incoherent response shows a temperature dependence linked to the onset of the CDW phase transition. Specifically, the lifetime of the first fast decay increases linearly with temperature until it reaches T$_{\text{CDW}}$, then it undergoes a rapid drop \cite{Tuniz2023}. It is worth noting that the lifetime of such a decay process does not show any divergence, contrary to what was reported for other Kagome systems~\cite{Zhao_2021}. On top of this incoherent response, marked coherent oscillations linked to the excitation of the amplitude mode (AM) of the CDW phase are detected \cite{Schaefer2013,Pokharel2022,Tuniz_2023PRR,Tuniz2023}. From these oscillations, the frequency of the phonon mode can be extracted (we refer the reader to the Supplementary Information for details about the methodology), providing information about the evolution of the phonon mode belonging to the flat phonon band as a function of temperature and strain. As expected from the AM of the system, we observe a progressive softening of the frequency of the mode upon approaching the CDW transition temperature, equal to 92 K (Fig.~\ref{figure2}(b)-(c)). This behavior, together with the increased damping, has been used to identify the AM of the system in this compound~\cite{Tuniz2023}, and in several other materials, such as several transition metal dichalcogenides~\cite{Demsar_1999,Vorobeva_2011,Tuniz_2023PRR,Schafer_2010}. 

Upon application of uniaxial strain (see the Supplementary Information for the full TR-OS datasets acquired in the various strain configurations), the frequency of the phonon mode coupled to the CDW is significantly affected. We highlight that the straining procedure was monitored several times and excellent repeatability was achieved. The results reported in Fig.~\ref{figure2}(c) demonstrate that, at low temperature, the frequencies are only weakly affected, since at about T=20 K no difference is observed outside the error bars. However, on approaching T$_{\text{CDW}}$, increasing strain leads to larger frequencies, as evidenced by the experimental points in Fig.~\ref{figure2}(c). In fact, going from the unstrained configuration (grey curve) to the largest applied strain (purple curve), the frequency changes from approximately 1.34 THz to 1.39 THz at 70 K. In other words, the effect of strain is that of enhancing the CDW in ScV$_6$Sn$_6$, that results in a weaker phonon softening, as signalled by the larger frequency of the amplitude mode. In the vicinity of the transition temperature, strain application has a marked effect on the phonon frequency, which competes with the temperature-induced phonon softening in the investigated temperature range. The strain-induced reduction of the AM phonon softening, more pronounced close to Tc, where the order parameter of the CDW is weaker, signals that strain application can effectively enhance the CDW in ScV$_6$Sn$_6$. Remarkably, our measurements bring evidence of the robustness of the CDW, and suggest a protocol to enhance it, through targeted strain-control. To better understand this behaviour, and to quantitatively estimate the strength of the applied strain, we performed extensive DFT calculations in the CDW phase, where the effect of strain has been considered. Note that we also estimated the strain from quantification of the mechanical deformation on the plate, as also described in Ref. \cite{LiuStrain} and shown in the Supplementary Information for completeness and comparison to previous works. However, the actual strain transferred to the sample is likely less than the one obtained by using this crude approach, whose upper limit is found to give $2.4\%$, which is $25‰$ larger than the value estimated from DFT.
\begin{figure*}
\centering
\includegraphics[width=\textwidth,angle=0,clip=true]{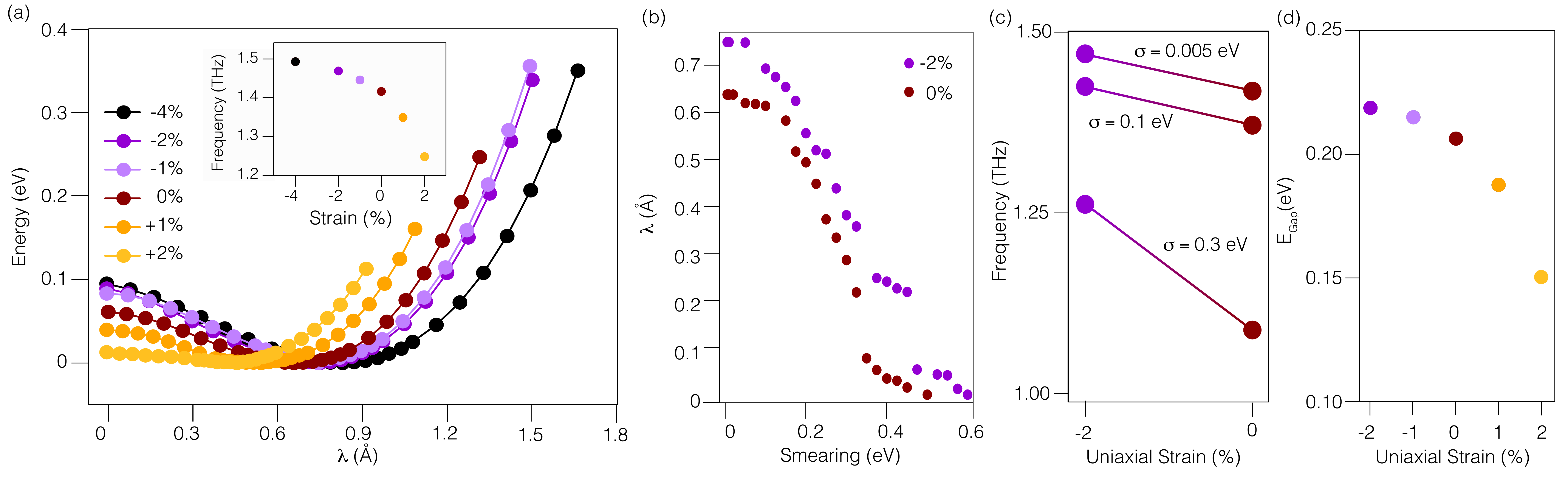}
\caption{(a) Total energy of ScV$_6$Sn$_6$ as a function of the polarization length $\lambda$. $\lambda = 0$ corresponds to the pristine phase. The inset contains the frequency values computed from a quadratic interpolation around the energy minimum. All the calculations in this panel have been performed with a smearing $\sigma =$ 0.005 eV. (b) Polarization length $\lambda$ as a function of the electronic smearing $\sigma$, for two uni-axial strain values (0\% and -2\%). (c) Phonon's frequency as a function of the uni-axial strain, for different smearing values. Destabilized CDW structures show more pronounced variations as a function of strain. (d) Energy gap opened by the CDW, computed midway between the A and L high-symmetry points, as a function of the applied uni-axial strain. Note the same trend of the inset in Fig. \ref{figure3}(a).}
\label{figure3}
\end{figure*}
DFT calculations were performed using VASP~\cite{Kresse_1996, PhysRevB.59.1758}, employing the projector augmented wave (PAW) method \cite{PhysRevB.50.17953}. Exchange and correlation effects have been included via the generalized gradient approximation (GGA) \cite{PhysRevA.38.3098} with the approach of Perdew-Burke-Ernzerhof (PBE)  \cite{PhysRevLett.77.3865}. A plane-wave cutoff of 500 eV and a 12$\times$12$\times$12 $\Gamma$-centered $k$-mesh were employed for accurate sampling of the Brillouin zone. Ionic relaxations were performed under constant volume conditions, while uni-axial strain was imposed by adjusting the lattice parameters, particularly the $a$-axis, and correspondingly straining or compressing the $c$-axis.
Convergence criteria were set for both ionic and electronic relaxations, with thresholds of 10$^{-5}$ eV/\AA\, and 10$^{-10}$ eV for forces and total energy, respectively.  The focus is on the influence of smearing and strain on the charge-density wave (CDW) phase. For each smearing value, the amplitude of the phonon mode between the CDW and pristine phases was computed, alongside the analysis of phonon frequencies.
These were determined by fitting a quadratic function around the minimum of the Born-Oppenheimer potential corresponding to the CDW phase, allowing to determine the ``spring" constant $k$. The effective mass m$^*$ was computed using the normalized displacement vector and the mass tensor. Finally, the phonon frequency $\nu$ was computed as $\nu = \omega/(2\pi)$, where $\omega = \sqrt{k/m^*}$. Spin-orbit coupling effects were not considered in this set of calculations. As in our previous work \cite{Tuniz2023}, electronic smearing is used to destabilize the strength of the CDW instability by simulating the effects of an electronic temperature, averaging over the states around the Fermi level~\cite{Marzari}.
To understand the effect of mechanical deformations and temperature variations, we compute via DFT the expected phonon's frequency for several values of uni-axial strain and electronic smearing $\sigma$ (Fig.~\ref{figure3}). The polarization length $\lambda$ of the material, defined as $\lambda = |\mathbf{r}_{\text{non-pristine}} - \mathbf{r}_{\text{pristine}}|$, is a function of both physical quantities.  The total energy is analyzed as a function of $\lambda$, where the pristine phase corresponds to $\lambda = 0$, and increasing values of $\lambda$ denote an interpolation toward and beyond the CDW phase (see Fig.~\ref{figure3}(a)). For every strain value, the minimum in energy is achieved when $\mathbf{r}_{\text{non-pristine}} \sim \mathbf{r}_{\text{CDW}}$. Owing to volume conservation, imposed modifying the $c$-axis length, the point of energy minimum shifts toward larger polarization lengths.
We further observe that $\lambda_{\text{max}}$ increases as well, while applying larger compressive uni-axial strain values. Finally, we note that increased strain values lead to a more pronounced energy difference between the pristine and CDW phases. Consequently, the frequency values computed from a quadratic interpolation around the energy minimum reveal a dependence on the applied strain (see inset in Fig. \ref{figure3}(a)), with larger compressive strains corresponding to larger frequencies. On the other hand, increased smearing values tend to destabilize the CDW phase. It can be observed from Fig. \ref{figure3}(b) the trend of $\lambda$ as a function of $\sigma$. Each data point has been computed following an ionic relaxation with the chosen smearing value. The polarization length for the -2\% case is always bigger than the corresponding 0\% case, since the former $c$-axis’ length is larger than the latter. While increasing the smearing value, the system moves toward a destabilized CDW configuration, hence decreasing $\lambda$. Additionally, to elucidate the relation between strain and electronic temperature, we examine the phonon frequency as a function of uni-axial strain for various $\sigma$ values (Fig. \ref{figure3}(c)). Notably, CDW structures destabilized by large smearing values, as large as 0.3 eV, exhibit more pronounced variations with strain than the small-smearing-counterpart. Lastly the energy gap $E_{gap}$, opened by the CDW and computed halfway between the A and L high-symmetry points (Fig. \ref{figure1}(e)), is analyzed as a function of strain (Fig. \ref{figure3}(d)). In line with the findings illustrated in Fig. \ref{figure3}(a), we observe a strikingly similar trend occurring also in Fig. \ref{figure3}(d), emphasizing the interplay between strain and polarization length within this material's CDW phase.

The calculated frequencies are in agreement with the data, both qualitatively and quantitatively, allowing us also to attribute, as already mentioned above, a level of strain as large as -0.5\% and -1.8\%. Note that such amount of strain is very large; not only it does not destabilize the CDW, thus corroborating the robustness of this phase transition, but it enhances the CDW, as signalled by the behavior of the CDW gap reported in Fig. \ref{figure3}(d).

In conclusion, we show a methodology based on application of uniaxial strain to control the frequency of the phonon mode which drives the unconventional CDW in ScV$_6$Sn$_6$. We demonstrated that the CDW is robust against the applied strain, with the effect of being enhanced. This mechanism is also intertwined to the temperature. In particular, the lower the temperature, the smaller the effect of strain on the CDW. On the other hand, on approaching the transition temperature, the same amount of strain markedly alters the phonon frequencies. Acting on the energy of phonon modes might be important in the context of electron-phonon coupling, being the Eliashberg function \cite{Ruf_2021, Liu_2022Str} dependent on both electron density of states and phonon density of states, and our study motivates future investigations aimed at understanding the effect of strain on the system's electronic structure.
%Here, we benchmarked the stability and robustness of the unconventional CDW in ScV$_6$Sn$_6$ and we unveiled a reliable strategy to enhance it based upon strain control.

\textit{Acknowledgements --} F.M. greatly acknowledges the SoE action of pnrr, number SOE\_0000068. A. C. acknowledges support from PNRR MUR project PE0000023-NQSTI. A.C., R.T., G.S. and D.D.S. acknowledge the Gauss Centre for Supercomputing e.V. (https://www.gauss-centre.eu) for funding this project by providing computing time on the GCS Supercomputer SuperMUC-NG at Leibniz Supercomputing Centre (https://www.lrz.de).
The research leading to these results has received funding from the European Union's Horizon 2020 research and innovation programme under the Marie Sk{\l}odowska-Curie Grant Agreement No. 897276. We are grateful for funding support from the Deutsche Forschungsgemeinschaft (DFG, German Research Foundation) under Germany’s Excellence Strategy through the Würzburg-Dresden Cluster of Excellence on Complexity and Topology in Quantum Matter ct.qmat (EXC 2147, Project ID 390858490) as well as through the Collaborative Research Center SFB 1170 ToCoTronics (Project ID 258499086).

%\end{scriptsize}

%\begin{scriptsize}
\bibliographystyle{naturemag}
\bibliography{main.bib}

\end{document}